\documentstyle[prl,aps]{revtex}

\begin{document}   
\twocolumn
\title{Recent rigorous results support the predictions of 
spontaneously broken replica  symmetry for realistic spin glasses} 
\author{Giorgio Parisi}
\address{Dipartimento di Fisica, Universit\`a di Roma {\sl La  Sapienza}\\ 
ed INFN Sezione di Roma I \\ 
Piazzale Aldo Moro, 00185 Roma (Italy)}
\maketitle

\date{\today}

\maketitle


\begin{abstract}
We show that the predictions of  spontaneously broken
 replica 
symmetry  are in 
perfect agreement with  the recent rigorous results 
obtained by Newman and Stein.
\end{abstract} 


In a very interesting paper Newman and 
Stein \cite{NS} have obtained new exact 
results for realistic spin glasses.  In this note I observe that their 
results for the quantities that they define are in perfect agreement with 
the predictions of the spontaneously broken
 replica 
symmetry (SBRS). 
For the reader convenience I recall some predictions  
of SBRS.

{\bf 1.} We consider a spin glass model at low temperature.  The system is 
in a box of side $L$, volume $V=L^D$, with fixed boundary condition (e.g.  
periodic).  The Hamiltonian depends on a set of quenched variables $J$.  We 
define the overlap $q$ as $V^{-1}\sum_i \sigma_i \tau_i$, where $\sigma$ 
and $\tau$ are two equilibrium configurations.  The probability 
distribution of $q$ is $P_J(q)$.  Denoting by an upper bar the average over 
the quenched disorder $J$ we can define the function
$P(q_1)  \equiv \overline{P_J(q_1)}$.

In the spin glass phase the funtion is not a simple delta function; it 
has a more complex structure, which signals the existence of many 
equilibrium states. 

The same results are obtained if before taking the $L\to \infty$ limit and 
averaging over the quenched disorder we fix to arbitrary values the 
couplings $J$ in a region of finite size, i.e.  we only average over the 
couplings not belonging to such a finite region.

{\bf 2.}
If the function $P(q_1)$ is not a delta function, one finds that the 
function $P_J(q_1)$ does depend on $J$ and it is not a self-averaging 
quantity. In other words the quantity $ P(q_1,q_2) \equiv
\overline{P_J(q_1),P_J(q_2)}$
is not given by the product.$P(q_1)P(q_2)$.  (Rigorous arguments in this 
direction  have been recently given by Guerra\cite{GU}). 

{\bf 3.} If we consider two copies of the same system, which have a mutual 
overlap $q$ in the $L \to \infty$ limit, the probability of finding a 
region of size $R$ where the overlap is $p\ne q$ goes to zero as $\exp( 
-R^\alpha f(p,q))$.  The authors of \cite{FPV} have estimated the exponent 
$\alpha$ to be equal to $D-\frac52$.

{\bf 4.} If the spins $\sigma$ and $\tau$ belong to two systems which 
differ in the value of the quenched disordered couplings $J$ in a finite 
(arbitrarily small) portion of the lattice links \cite{K}, they turn out to 
have an overlap which is always equal to the minimum allowed ($0$ in zero 
magnetic field). Spin glasses have a chaotic dependence 
on the coupling constant.
     
An apparent contradiction with SBRS is detected in the recent paper 
\cite{NS}, in which the authors give two new definitions of a probability 
distribution of the overlaps $q$ (which we indicate, with abuse of 
language, again by $P_J(q)$).  Such $P_J(q)$ do not depend on $J$ in the 
large volume limit.  However the objects the authors define are different 
from the ones we usually encounter in the literature.  We will show in the 
following that this $J$-independence turns out to be in perfect agreement 
with the SBRS approach.

Let us 
see which are the predictions of the mean field theory for quantities that 
are defined in \cite{NS}.  For simplicity we will consider only two cases.

{\bf 1.} We consider a system of size $L$ and we concentrate our attention 
on what happens in a box of size $R$.  We define by $q_R$ the overlap of two 
replicas in this box.  We call $I$ the couplings inside the box and  
$E$ those outside the box.  
The couplings $J$ are obviously determined by $E$ and 
$I$ ($J=I\oplus E$) .
 	
Following the first construction of ref. \cite{NS} we define 
 	
\begin{equation}
  P^{(1)}_I (q) \equiv \int d\mu(E) P_J^R(q)\ ,
\end{equation}
where $ P_J^R(q)$ is the probability distribution of the overlap $q_R$, i.e. 
of the overlap restricted to the region $R$. 
 	
Let us first send $L \to \infty$, or if we prefer, let us consider the case 
$L>>R$.  When $R$ goes to infinity $q$ and $q_R$ are equal and $P^{(1)}_I 
(q)$ coincides with $\int d\mu(E) P_J(q)$.  This last integral does not 
depend on $I$, so for large $R$ mean field theory predicts that
 	
\begin{equation}
  P^{(1)}_I=P(q)\ ,
\end{equation}
and it is independent from $I$ as rigorously proven in \cite{NS}.
 	
{\bf 2.} Let us discuss a second definition, $P^{(2)}(q)$, which \cite{NS2} 
is inspired by the second construction of ref.  \cite{NS}.  We consider two 
systems, one with couplings $J_1=I\oplus E_1$ and the other with couplings 
$J_2=I\oplus E_2$.  We consider the distribution probability of the 
overlaps $q_R$ and $q$ among a configuration of the first system and a 
configuration of the second system, the first overlap ($q_R$) being 
restricted to the region of size $R$ (where the couplings are $I$ for both 
systems).  We introduce the corresponding probability distributions which 
obviously  depend on the couplings $I$, $E_1$ and $E_2$.
 	
We define 
\begin{eqnarray}\nonumber
  P^{(2)}_I (q_R) 
  \equiv \int d\mu(E_1)  d\mu(E_2) P_{I,E_1,E_2}^R(q_R)\ , \\
  P_I (q) \equiv\int d\mu(E_1)  d\mu(E_2) P_{I,E_1,E_2}(q) \ .
\end{eqnarray}
Also in this case $P_I (q)$ and $P^{(2)}_I (q)$ coincide in the large $R$ 
limit. $P^{(2)}_I (q) =\delta(q)$, due to the chaotic nature of spin glasses.
$ P^{(2)}_I$ is independent from $I$, as proven in 
\cite{NS} and it is different from $ P^{(1)}_I$.
 
The SBRS theory is sophisticated enough to give different answers to 
different questions, spurring from different definitions of the overlap 
distribution.  In all cases it gives the correct answer, i.e.  
it declares self-averaging objects that one can rigorously prove to be 
self-averaging (even for realistic spin glasses) and non-self-averaging 
quantities that are rigorously shown to be 
non-self-averaging.
 
It is a pleasure for me to thank C. Newman and D. Stein for an illuminating 
correspondence, and F. Guerra and E. Marinari for very interesting 
discussions.

\end{document}